\documentclass[intlimits,twoside,a4paper]{article}

\usepackage{amsmath,amssymb}
\usepackage{graphicx}
\usepackage{wrapfig}

\usepackage[T2A]{fontenc}
\usepackage[cp1251]{inputenc}

\usepackage{epsf,multicol,ifthen}

\usepackage[eqsecnum]{cmpj2}

\issue{2015}{18}{4}{43801}
\doinumber{10.5488/CMP.18.43801}

\title[Temperature regimes of formation of nanometer periodic structure of adsorbed atoms in GaAs]%
{Temperature regimes of formation of nanometer periodic structure of adsorbed atoms in GaAs semiconductors under the action of laser irradiation}
\author[R.M.~Peleshchak, O.V.~Kuzyk, O.O.~Dan'kiv]
{R.M.~Peleshchak\thanks{E-mail: peleshchak@rambler.ru}\,, O.V.~Kuzyk, O.O.~Dan'kiv}
\address{Drohobych Ivan Franko State Pedagogical University,
24~Franko St., 82100 Drohobych, Ukraine}

\date{Received October 4, 2015, in final form November 15, 2015}
\authorcopyright{R.M.~Peleshchak, O.V.~Kuzyk, O.O.~Dan'kiv, 2015}

\begin{document}

\maketitle

\begin{abstract}
The theory of nucleation of nanoscale structures of the adsorbed atoms (adatoms), which occurs as a result of the self-consistent interaction of adatoms with the surface acoustic wave and electronic subsystem is developed. Temperature regimes of formation of nanoclusters on ${n}$-GaAs surface under the action of laser irradiation are investigated. The offered model permits to choose optimal technological parameters (temperature, doping degree, intensity of laser irradiation) for the formation of the surface periodic defect-deformation structures under the action of laser irradiation.
\pacs 81.07.Bc, 66.30.Lw
\end{abstract}

\section{Introduction}

Recently, we have observed intensive researches into obtaining semiconductor structures with self-assembled nanoclusters by methods of molecular beam epitaxy \cite{Zin15,Ipa98}, ion implantation \cite{Yam03,Bor12} and under the action of laser irradiation \cite{Pel09,Vla15,Vel10}, as well as the possibility of controlling their physical properties.

The non-linear diffusion-deformation theory of self-organization of nanoclusters of implanted impurities in semiconductors that considers an elastic interaction of the implanted impurities among themselves and with atoms of a matrix was developed in \cite{Pel13}. The impurity which gets to a matrix, leads to changes in its volume and energy, and the initial fluctuation of deformation under certain conditions causes the emergence of deformation-induced flows of the implanted impurities. As a result, there are forces in the non-uniform deformation-concentration field  proportional to gradients of concentration and deformation which further deform the matrix. These forces are caused by an increase of initial fluctuation and lead to self-organization of clusters of impurities.

The information on nucleation (incipient state of the formation) of periodic nanostructures of ad\-atoms and the implanted impurity is important for optimization of technological process and the predicted controlling of physical parameters of semiconductor structures with nanoclusters \cite{Sle09,Sug09,Aar10,Lvo15,Dub11}. In particular, to calculate the symmetry, the period and the formation time of surface structures it is sufficient to analyze only the initial (linear) stage of the development of the defect-deformation instability \cite{Eme008}.

The theory of spontaneous nucleation of the surface nanometer lattice which is caused by instability in the system of the adatoms interacting with the self-consistent surface acoustic wave (SAW) was developed in \cite{Eme008, Eme08}. Within this theory, the conditions of the formation of nanoclusters on the surface of solid states are established and the periods of a nanometer lattice as functions of concentration of adatoms and temperature are defined. However, the offered theory adequately describes the processes of the formation of nanoclusters only at low temperatures. It is based on the fact that this model does not consider the temperature dependence of concentration of the adatoms, as well as the interaction of electronic and defect subsystems that significantly depends on the temperature.

Due to the high mobility of charge carriers, gallium arsenide is widely used in the production of quantum-dimensional structures and high-frequency lasers on their basis. A considerable attention of researchers has recently been paid to laser modification of the morphology of near-surface GaAs layers at which the conditions of the formation of nanoclusters are controlled \cite{Pel09,Vla15}. The purpose of this work is to study, on the initial states, the conditions of the formation of surface structures of adatoms and their periods in a semiconductor under the action of laser irradiation depending on temperature and doping degree. In this work, the temperature regimes and the influence of the degree of a doping of GaAs semiconductor on the formation of periodic nanometer structures of adatoms and their periods under the action of laser irradiation are investigated.

\section{The model}

The equations for the displacement vectors $\vec u$ of an elastic medium is of the form \cite{Lan70}:
\begin{equation}
\label{eq1}
\frac{{\partial ^2 \vec u}}{{\partial t^2 }} = c_\textrm{t}^2 \Delta \vec u + (c_\textrm{l}^2  - c_\textrm{t}^2 )\textrm{grad}\left( {\textrm{div}\,\vec u} \right),
\end{equation}
where $c_\textrm{l}$ and $c_\textrm{t}$ are longitudinal and transversal sound velocities, respectively.

Let the surface of semiconductor coincide with the plane $z = 0$ ($z$-axis is directed into the crystal depth). Let us assume that the surface perturbation of an elastic medium arises along the $x$-axis. We represent this perturbation in the form of a static SAW, which quickly fades into the depth of a semiconductor and has an amplitude growing with time \cite{Eme08}:
\begin{align}
\label{eq2}
u_x  &=  - \ri q R\re^{\ri qx + \lambda t - k_\textrm{l} z}  - \ri k_\textrm{t} Q\re^{\ri qx + \lambda t - k_\textrm{t} z},\\
%
\label{eq3}
u_z  &= k_\textrm{l} R\re^{\ri qx + \lambda t - k_\textrm{l} z}  + qQ\re^{\ri qx + \lambda t - k_\textrm{t} z},
\end{align}
where  $k_\textrm{l,t}^2  = q^2  + {{\lambda ^2 }}/{{c_\textrm{l,t}^2 }}$;  $\lambda$ is the increment of defect-deformation instability \cite{Eme08}; $R$ and $Q$ are SAW amplitudes.

Then, deformation $\varepsilon$   on the semiconductor surface ($z = 0$) is of the form
\begin{equation}
\label{eq4}
\varepsilon  = \frac{{\partial u_x }}{{\partial x}} + \frac{{\partial u_z }}{{\partial z}} =  - \frac{{\lambda ^2 }}{{c_\textrm{l}^2 }}R\re^{\ri qx + \lambda t}.
\end{equation}

Let us consider the alloyed $n$-GaAs semiconductor containing impurities (ionization donors, free electrons and adsorbed atoms of Ga which are also ionization donors). In this case, the electroneutrality condition can be presented as follows:
\begin{equation}
\label{eq5}
n_0 (T) = N_\textrm{d}^ +  \left( T \right) + N_0^ +  (T),
\end{equation}
where $N_\textrm{d}$, $n_0$ and $N_0$ are the surface concentration of ionized donors, the spatially homogeneous values of the surface concentration of electrons and adatoms, respectively.

Concentration of adatoms can be presented as the sum of concentration $N_{0r}^ +  \left( T \right)$  of equilibrium defects and concentration $N_{0l}^ +$  of the defects generated under the action of laser irradiation that depends on its power:
\begin{equation}
\label{eq6}
N_0^ +  (T) = N_{0r}^ +  (T) + N_{0l}^ +,
\end{equation}
where  $N_{0r}^ +  \left( T \right) = N_{0r}^\infty  \exp \left( { - {{\Delta U}}/{{k_\textrm{B} T}}} \right)$;  $\Delta U$ is the formation energy of an adatom of Ga;  $k_\textrm{B}$  is Boltzmann constant; $T$ is temperature;  $N_{0r}^\infty$ is the surface concentration of atoms of Ga.

The direction of the surface wave is determined by anisotropic properties of crystals. In particular, in \cite{Oht89} it is shown that the surface diffusion coefficient of Ga on GaAs (001) in the $[0\bar 11]$  direction is four times larger than the surface diffusion coefficient in the $\left[ {011} \right]$  direction.

Periodic surface deformation leads to spatial nonuniform redistribution of adatoms $N(x)$, the modulation of the bottom of the conduction band and, respectively, to redistribution of the concentration of conduction electrons $n(x)$ and the electrostatic potential  $\varphi(x)$:
\begin{align}
\label{eq7}
N(x) &= N_0  + N_1 (x) = N_0  + N_1 (q)\re^{\ri qx + \lambda t},\\
%
\label{eq8}
n(x) &= n_0  + n_1 (x) = n_0  + n_1 (q)\re^{\ri qx + \lambda t},\\
%
\label{eq9}
\varphi (x) &= \varphi (q)\re^{\ri qx + \lambda t},
\end{align}
where $N_1(q)$, $n_1(q)$, and  $\varphi(q)$ are the amplitudes of the corresponding periodic perturbations.

Then, taking into account (\ref{eq7}), (\ref{eq8}) and (\ref{eq9}), the Poisson's equation can be presented as follows:
\begin{equation}
\label{eq10}
 - q^2 \varphi (q) = \frac{e}{{\varepsilon _0 \tilde \varepsilon a}}\left[n_1 (q) - N_1 (q)\right],
\end{equation}
where  $\varepsilon_0$ and  $\tilde \varepsilon$ are dielectric constant and dielectric permittivity of the medium, respectively;  $a$ is the lattice constant.

The equations for concentration of the charged adsorbed atoms are of the form
\begin{equation}
\label{eq11}
\frac{{\partial N}}{{\partial t}} = D\frac{{\partial ^2 N}}{{\partial x^2 }} + \frac{\partial }{{\partial x}}\left({\mu N\frac{{\partial \varphi }}{{\partial x}}} \right) - D\frac{\theta }{{kT}}\frac{\partial }{{\partial x}}\left[ {N\frac{\partial }{{\partial x}}\left( {\varepsilon  + l_\textrm{d}^2 \frac{{\partial ^2 \varepsilon }}{{\partial x^2 }}} \right)} \right],
\end{equation}
where $D$ and $\mu$  are the surface diffusion coefficient and the mobility of adatoms which are related among themselves by Einstein's relation  $\mu  = D{e}/({{k_\textrm{B} T}})$;  $\theta$  is the deformation potential; $l_\textrm{d}$ is the characteristic length of interaction of adatoms with lattice atoms. The second term considers the interaction of adatoms with an electric field arising due to the spatial nonuniform redistribution of electric charge. The third term expresses the interaction of adatoms with the deformation field, taking into account the nonlocal interaction \cite{Eme08}.

Taking into account (\ref{eq4}), (\ref{eq7})--(\ref{eq10}) and in an approximation  $N_1<<N_0$, the equation (\ref{eq11}) can be written as follows:
\begin{equation}
\label{eq12}
\lambda N_1 (q) =  - Dq^2 N_1 (q) + \frac{{DN_0 \Phi }}{{kT}}\left[n_1 (q) - N_1 (q)\right] - \frac{{DN_0 \theta }}{{k_\textrm{B} T}}\left[ {\frac{{\lambda ^2 }}{{c_\textrm{l}^2 }}Rq^2 (1 - q^2 l_\textrm{d}^2 )} \right],
\end{equation}
where  $\Phi  = {{e^2 }}/({{\varepsilon _0 \tilde \varepsilon a}})$.

The density of the electron current
\begin{equation}
\label{eq13}
j = n\mu _n \frac{{\partial\chi }}{{\partial x}},
\end{equation}
where  $\mu_n(T, n)$ is the mobility of electrons which depends on the temperature and the doping degree of semiconductor \cite{Mna04}; the electrochemical potential $\chi$  is defined by  the relation
\begin{equation}
\label{eq14}
\chi (x) = k_\textrm{B} T\ln \frac{{n(x)}}{{N_i }} - e\varphi (x) + a_\textrm{c} \varepsilon (x),
\end{equation}
where $N_i$  is the effective density of states;
$N_i  = 2\left( {{{2\pi mk_\textrm{B}T}}/{{h^2 }}} \right)^{3/2}$; $a_\textrm{c}$ is the constant of hydrostatic deformation potential of the conduction band. Then, taking into account (\ref{eq13}) and (\ref{eq14}), the continuity equation can be presented as follows:
\begin{equation}
\label{eq15}
e\frac{{\partial n}}{{\partial t}} = k_\textrm{B} T\mu _n \frac{\partial }{{\partial x}}\left( {n\frac{\partial }{{\partial x}}\ln \frac{n}{{N_i }}} \right) - e\mu _n \frac{\partial }{{\partial x}}\left( {n\frac{{\partial \varphi }}{{\partial x}}} \right) + a_\textrm{c} \mu _n \frac{\partial }{{\partial x}}\left( {n\frac{{\partial \varepsilon }}{{\partial x}}} \right).
\end{equation}

Taking into account (\ref{eq4}) and (\ref{eq7})--(\ref{eq10}), the equation (\ref{eq15}) can be written as follows:
\begin{equation}
\label{eq16}
n_1 (q)\left( {e\lambda  + k_\textrm{B} T\mu _n q^2  + n_0 \mu _n \Phi } \right) = N_1 (q)n_0 \mu _n \Phi  + a_\textrm{c} n_0 \mu _n q^2 \frac{{\lambda ^2 }}{{c_\textrm{l}^2 }}R\,.
\end{equation}

Solving the system of equations (\ref{eq12}) and (\ref{eq16}), we obtain expressions for the amplitudes of the surface concentration of adatoms $N_1(q)$ and conduction electrons $n_1(q)$.

The spatial nonuniform distribution of adatoms modulates the surface energy $F(x)$, which leads to the appearance of lateral mechanical tension  $\sigma _{xz}  = {{\partial F\big(N(x)\big)}}/{{\partial x}}$, which is compensated by shift tension in the medium \cite{Eme08}. The boundary condition expressing the balance of lateral tension is as follows:
\begin{equation}
\label{eq17}
\frac{E}{{1 + \nu }}\left( {\frac{{\partial u_x }}{{\partial z}} + \frac{{\partial u_z }}{{\partial x}}} \right)_{z = 0}  = \frac{{\partial F\big(N(x)\big)}}{{\partial x}} = \frac{{\partial F}}{{\partial N}}\frac{{\partial N_1 (x)}}{{\partial x}},
\end{equation}
where $E$ and $\nu$ are Young's modulus and Poisson's ratio, respectively.

Besides, the interaction of adatoms with atoms of a semiconductor results in the emergence of the normal mechanical tension on the surface, and the corresponding boundary condition is of the form:
\begin{equation}
\label{eq18}
\frac{E}{{1 + \nu }}\left( {\frac{{\partial u_z }}{{\partial z}} + \frac{\nu }{{1 - 2\nu }}\frac{{\partial u_x }}{{\partial x}}} \right)_{z = 0}  = \frac{{\theta }}{a}N_1 (x).
\end{equation}

Thus, the system of homogeneous linear equations (\ref{eq17}) and (\ref{eq18}) for amplitudes $R$ and $Q$ is obtained and the dispersion dependencies  $\lambda(q)$ can be obtained from the condition of non-triviality of solutions (from the condition of equality to zero of the determinant of this system).

\section{Calculation results and their discussion}

The calculations of $\lambda(q)$ were carried out for the GaAs semiconductor doped by silicon at the following values of parameters: $l_\textrm{d} = 2.9$~nm; $a = 0.565$~nm; $c_{\,\text{l}} = 3500$~m/s;   $\nu= 0.318$;  $E = 0.85$~Mbar; $a_\textrm{c} = -7.17$~eV;  $\theta = 10$~eV;   $\tilde \varepsilon = 12$;
$D = 5 \times 10^7 \exp \left[ { - {{5.6}}/{{(k_\textrm{B} T)}}} \right]$~cm$^{2}$/s \cite{Bol72};  $N_{0r}^\infty = 3 \times 10^{14}$~cm$^{- 2}$;  $\Delta U = 0.1$~eV;   $N_\textrm{d}^ +   = N_{{0\textrm{Si}}} \exp[ - {{0.058}}/{{(k_\textrm{B} T)}}]$, where $N_{0\textrm{Si}}$ is concentration of Si impurities. Mobility of electrons as function of temperature and concentration of donors was determined by the technique given in work \cite{Mna04}.

\begin{figure}[!t]
\begin{center}
\begin{multicols}{2}
\includegraphics[width=70mm]{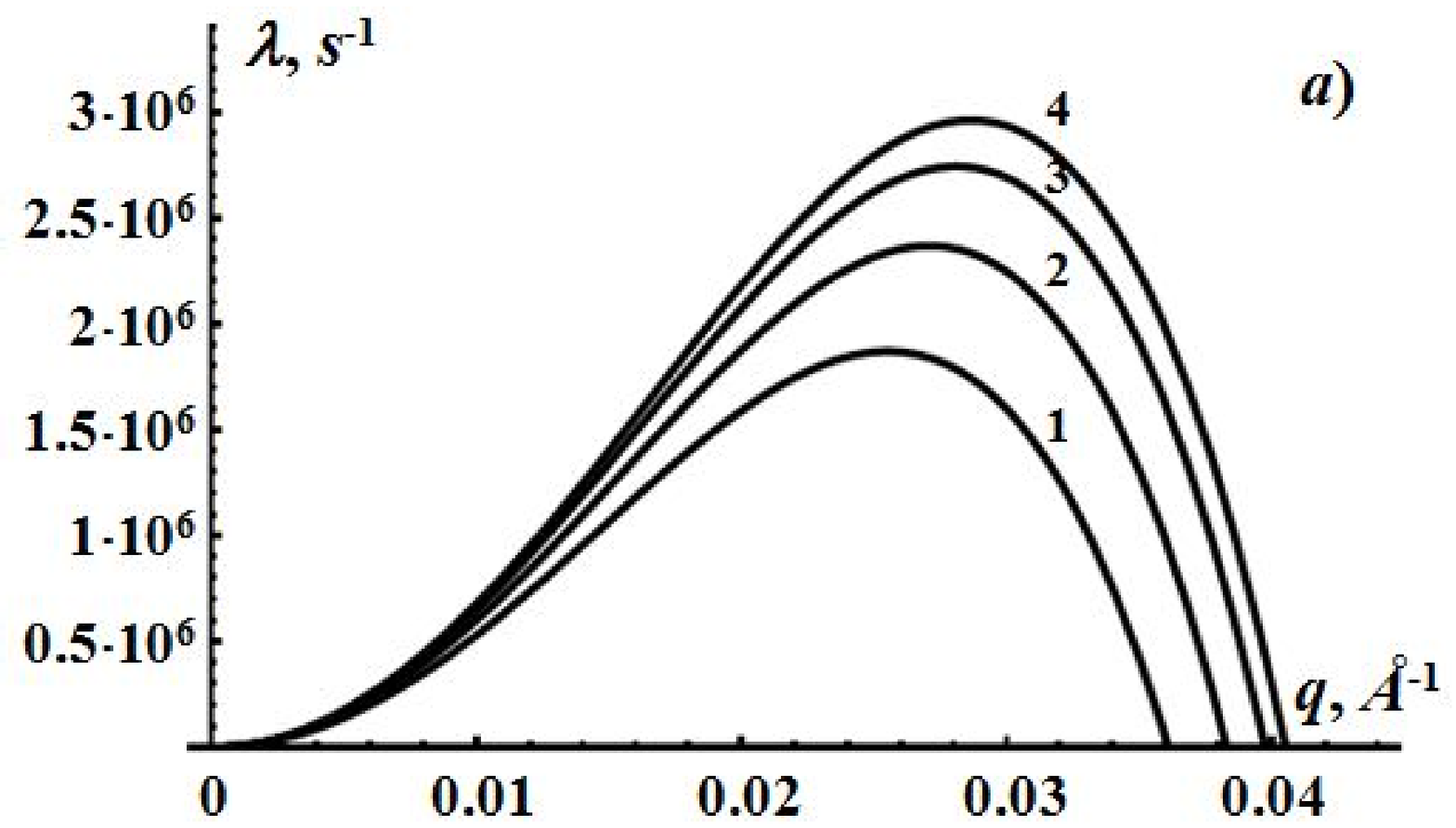}
\includegraphics[width=70mm]{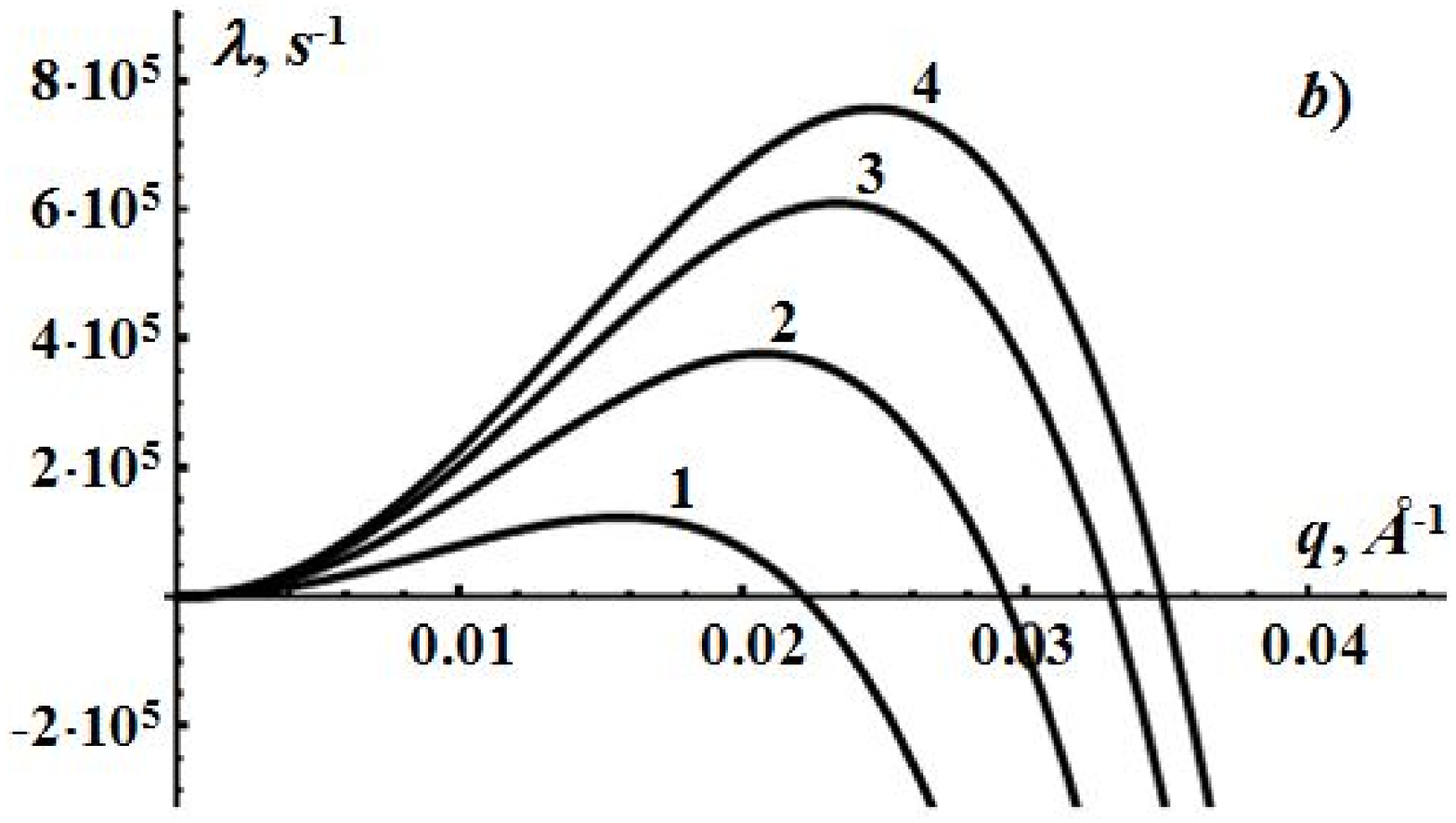}
\includegraphics[width=70mm]{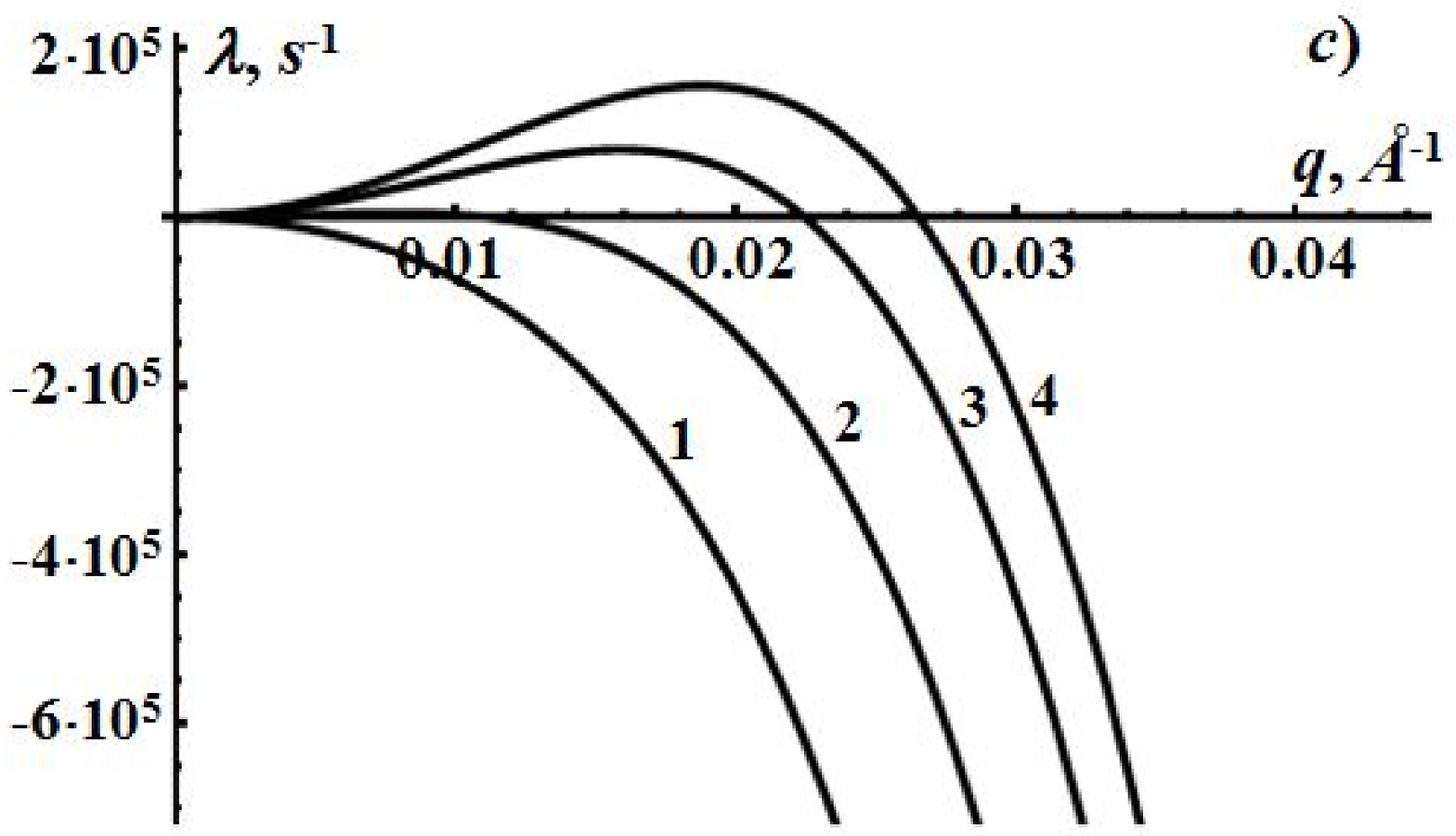}
\includegraphics[width=70mm]{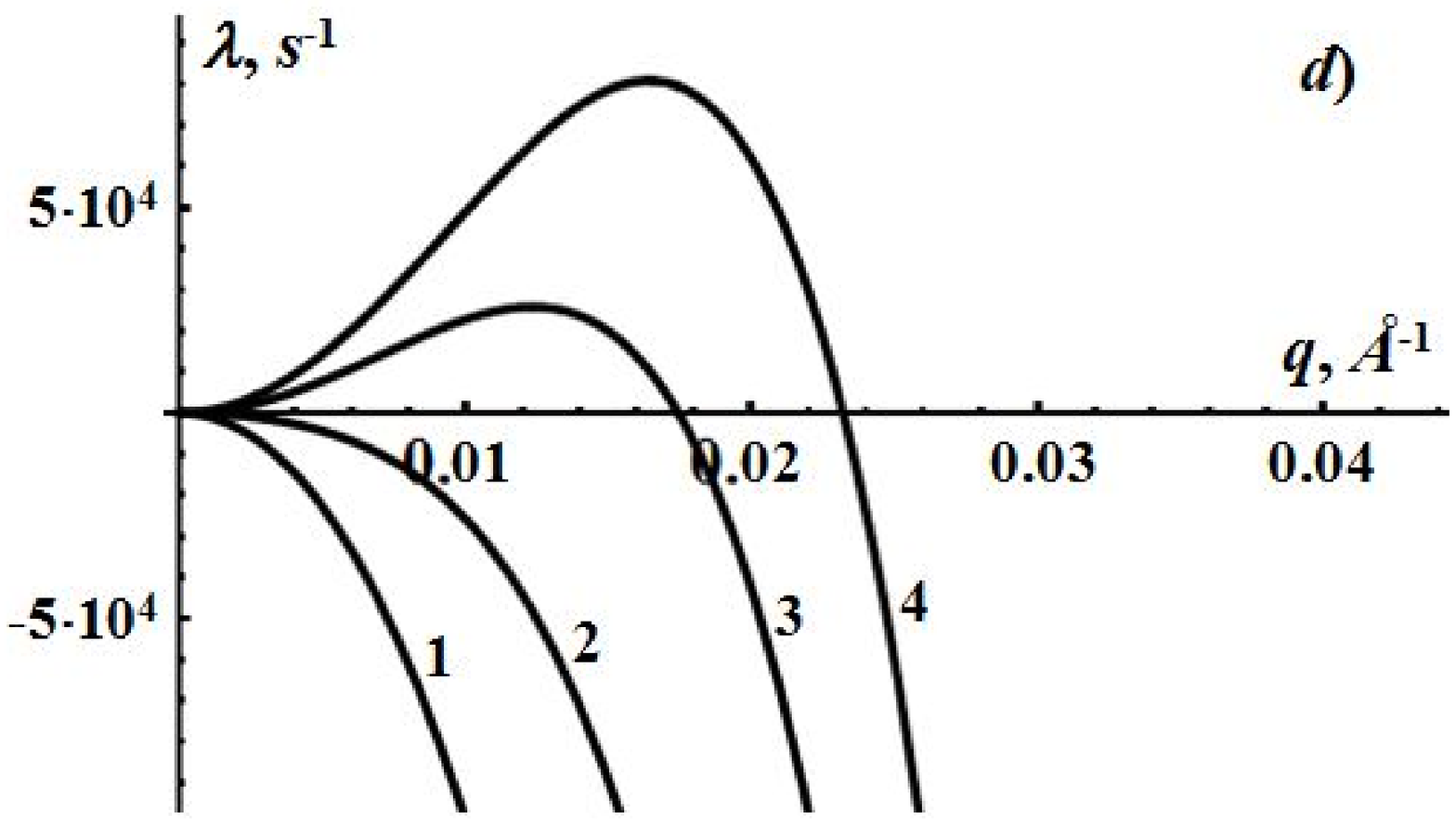}
\includegraphics[width=70mm]{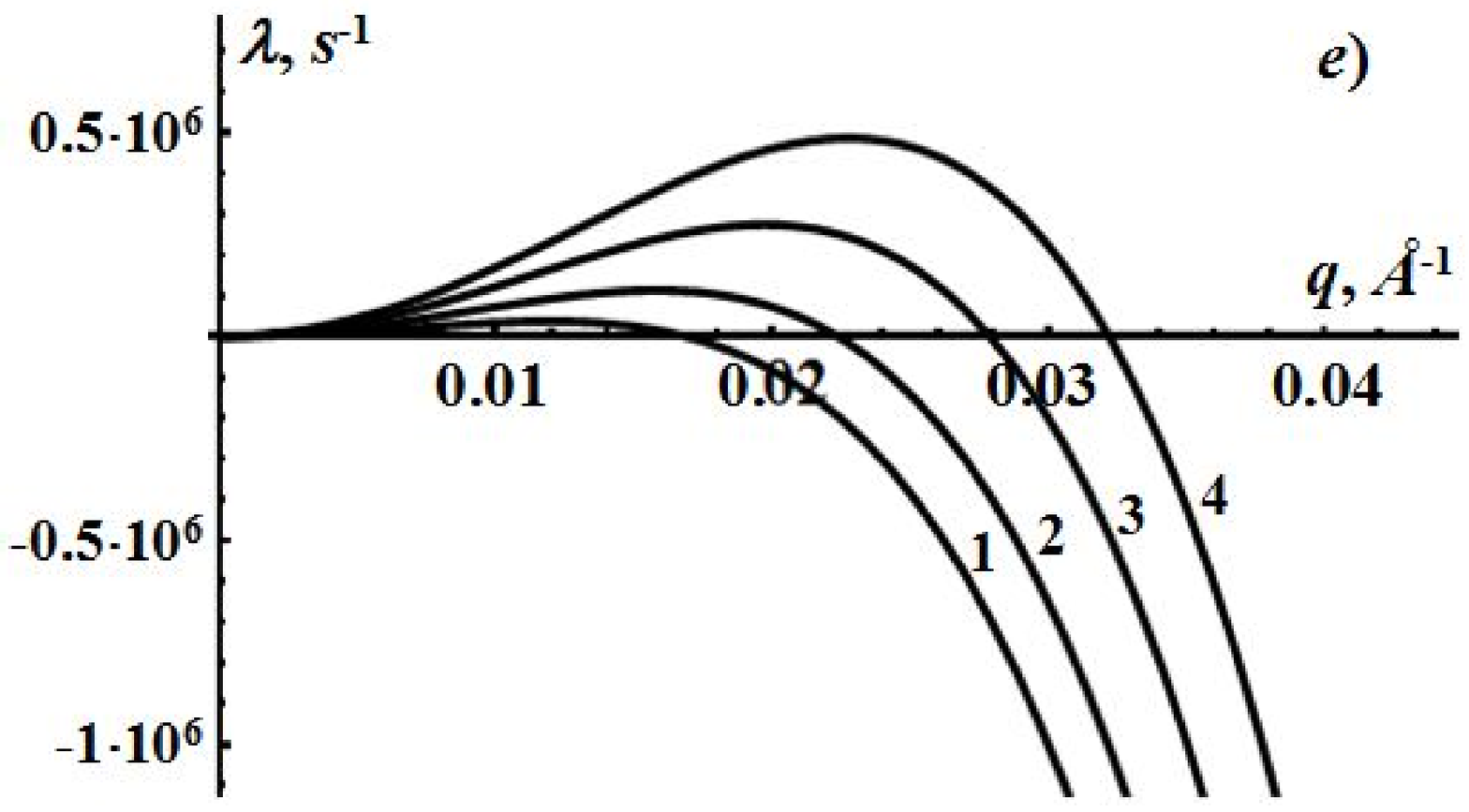}
\includegraphics[width=70mm]{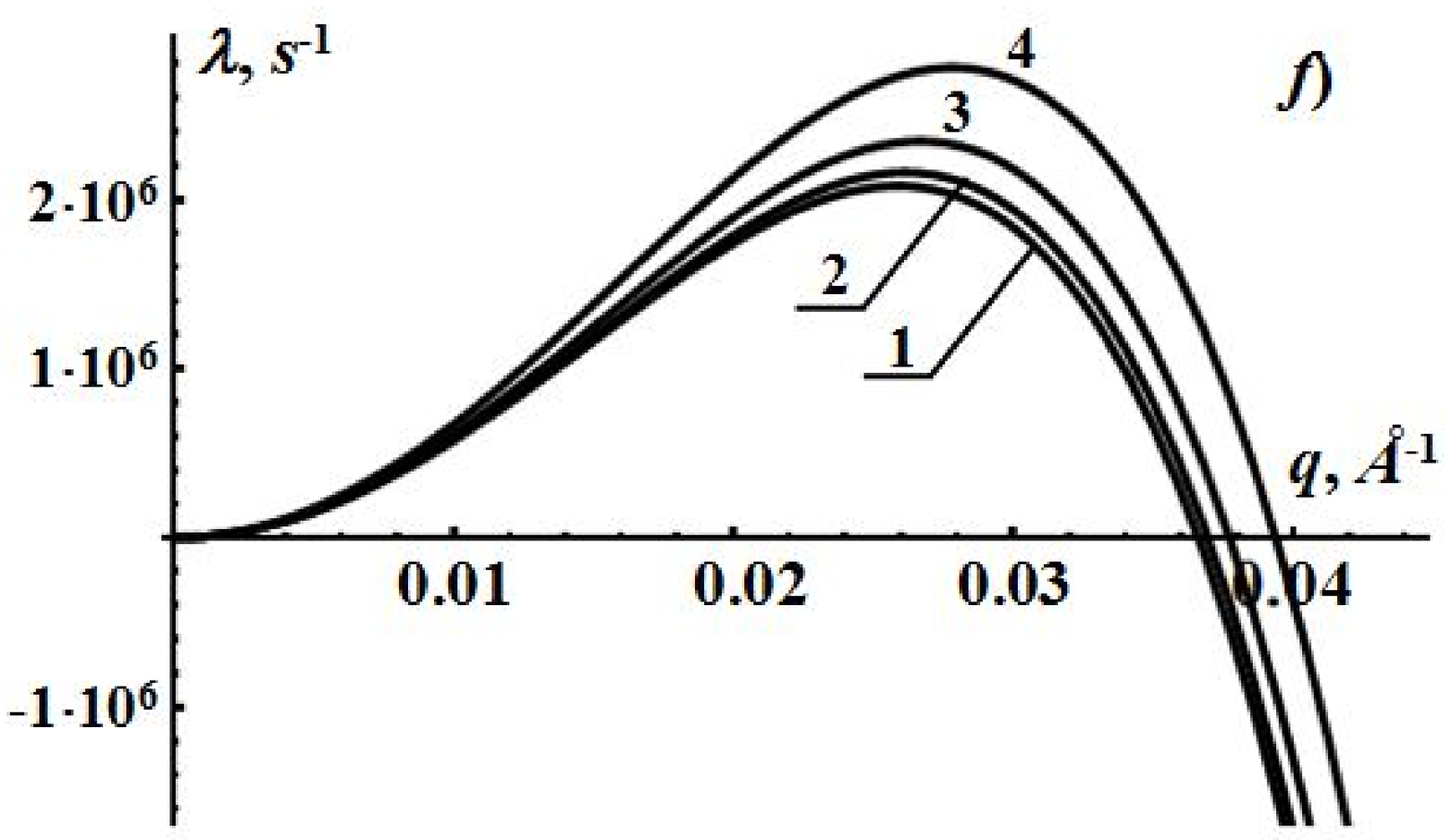}
\end{multicols}
\end{center}
\vspace{-3mm}
        \caption{The dependence of the increment of defect-deformation instability on the wave vector at concentration of the adsorbed atoms of
 $N_{0l}^ + = 5\times 10^{12}$~cm$^{-2}$ at various values of the concentration of donors:
1~--- $ N_{0\textrm{Si}} = 10^8$~cm$^{-2}$; 2~--- $ N_{0\textrm{Si}} = 5\times 10^{12}$~cm$^{-2}$; 3~--- $ N_{0\textrm{Si}} = 2\times10^{13}$~cm$^{-2}$; 4~--- $ N_{0\textrm{Si}} = 10^{14}$~cm$^{-2}$; (a)~$T = 70$~K; (b) $T = 150$~K; ({c}) $T = 250$~K; ({d}) $T = 300$~K; ({e}) $T = 500$~K; ({f}) $T = 800$~K.}
        \label{fig1}
    \end{figure}

\begin{figure}[!t]
\begin{center}
\begin{multicols}{2}
\includegraphics[width=70mm]{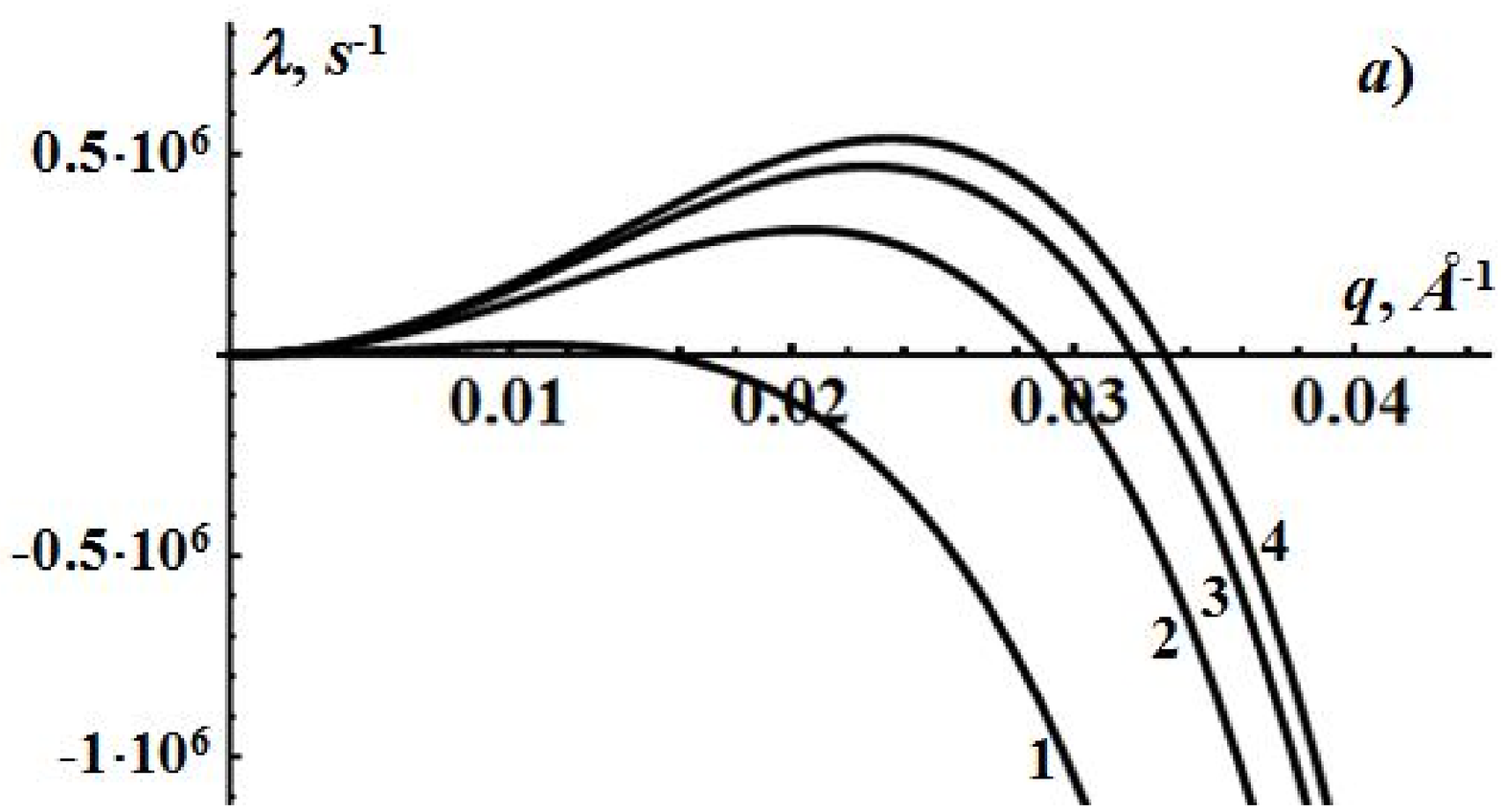}
\includegraphics[width=70mm]{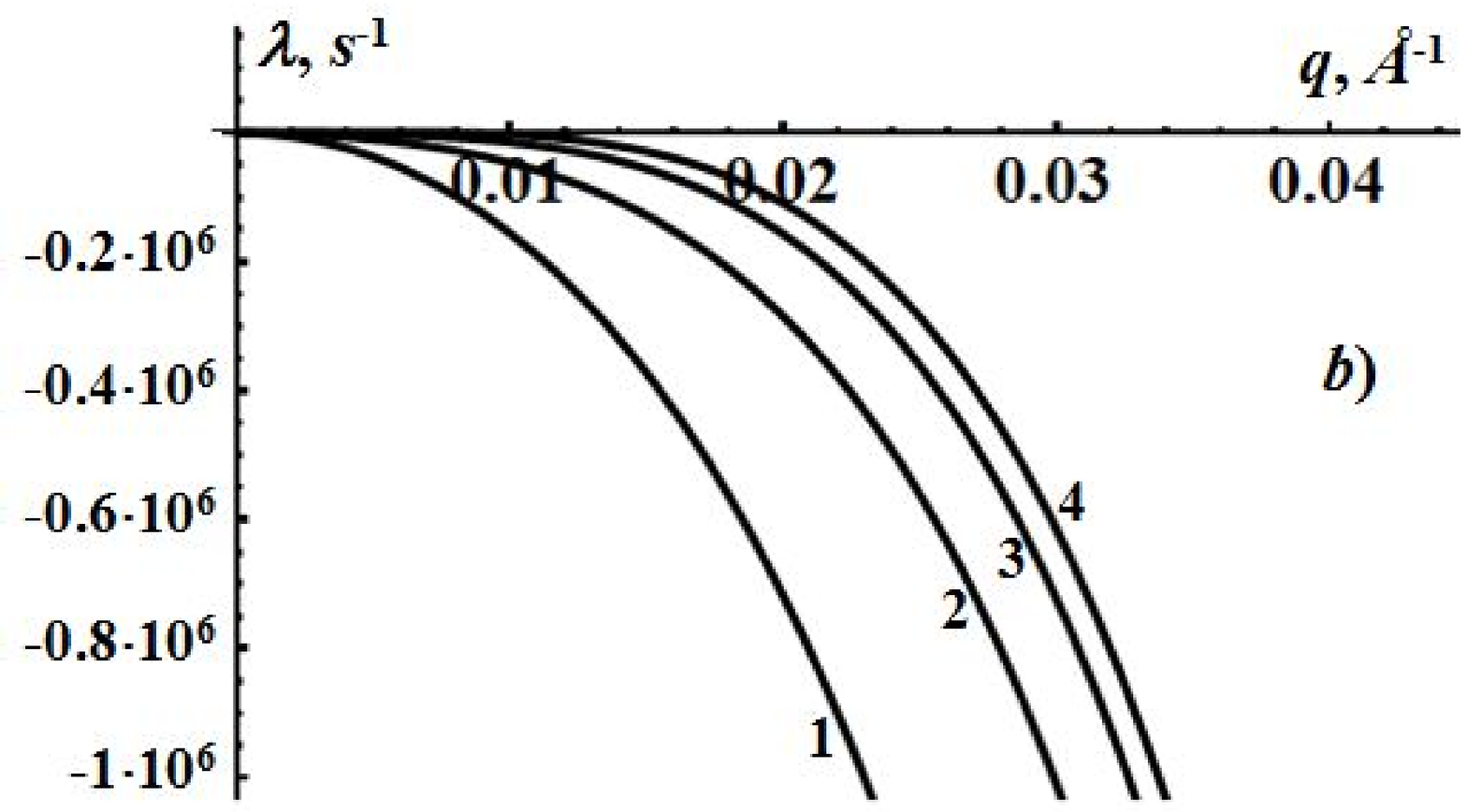}
\includegraphics[width=70mm]{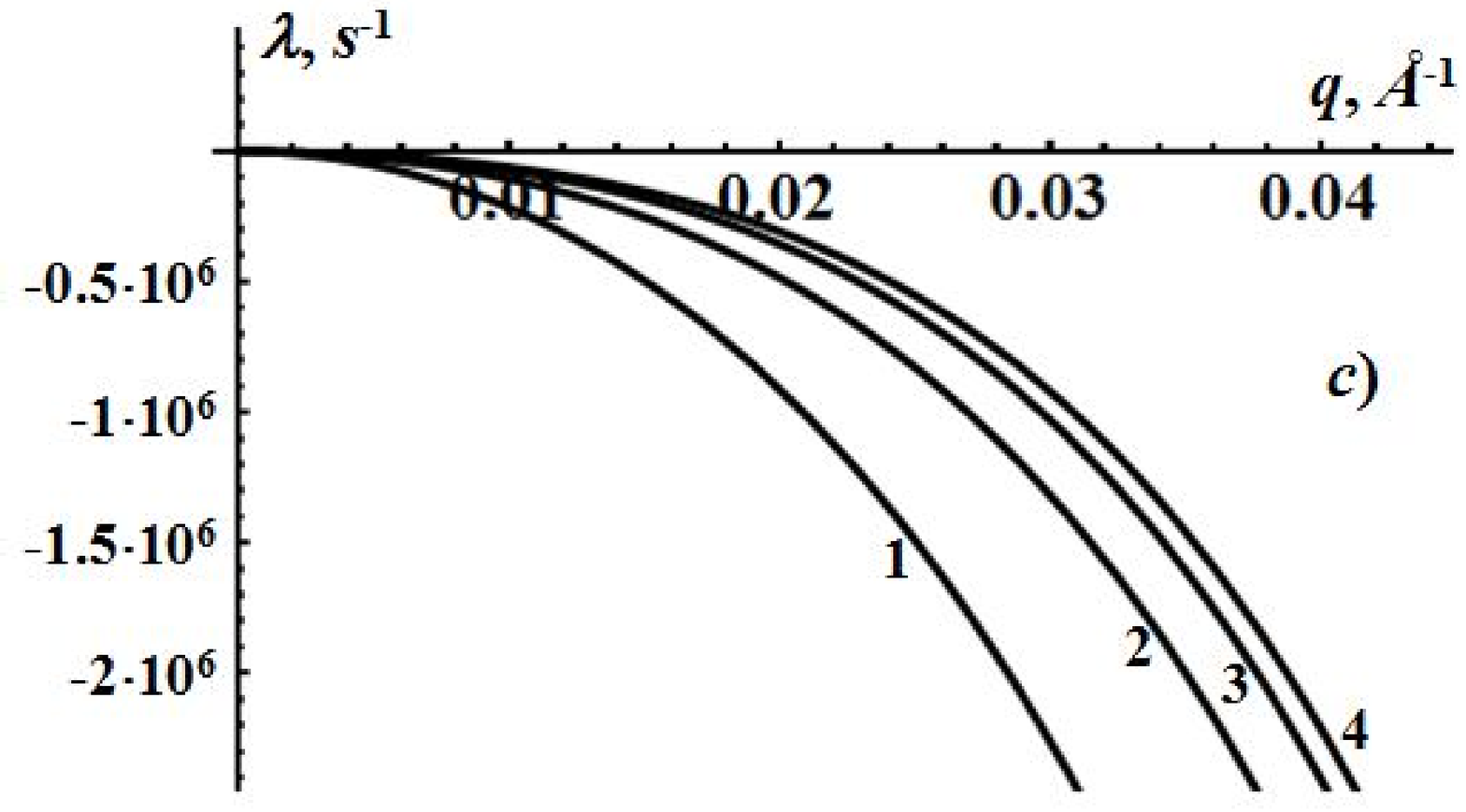}
\includegraphics[width=70mm]{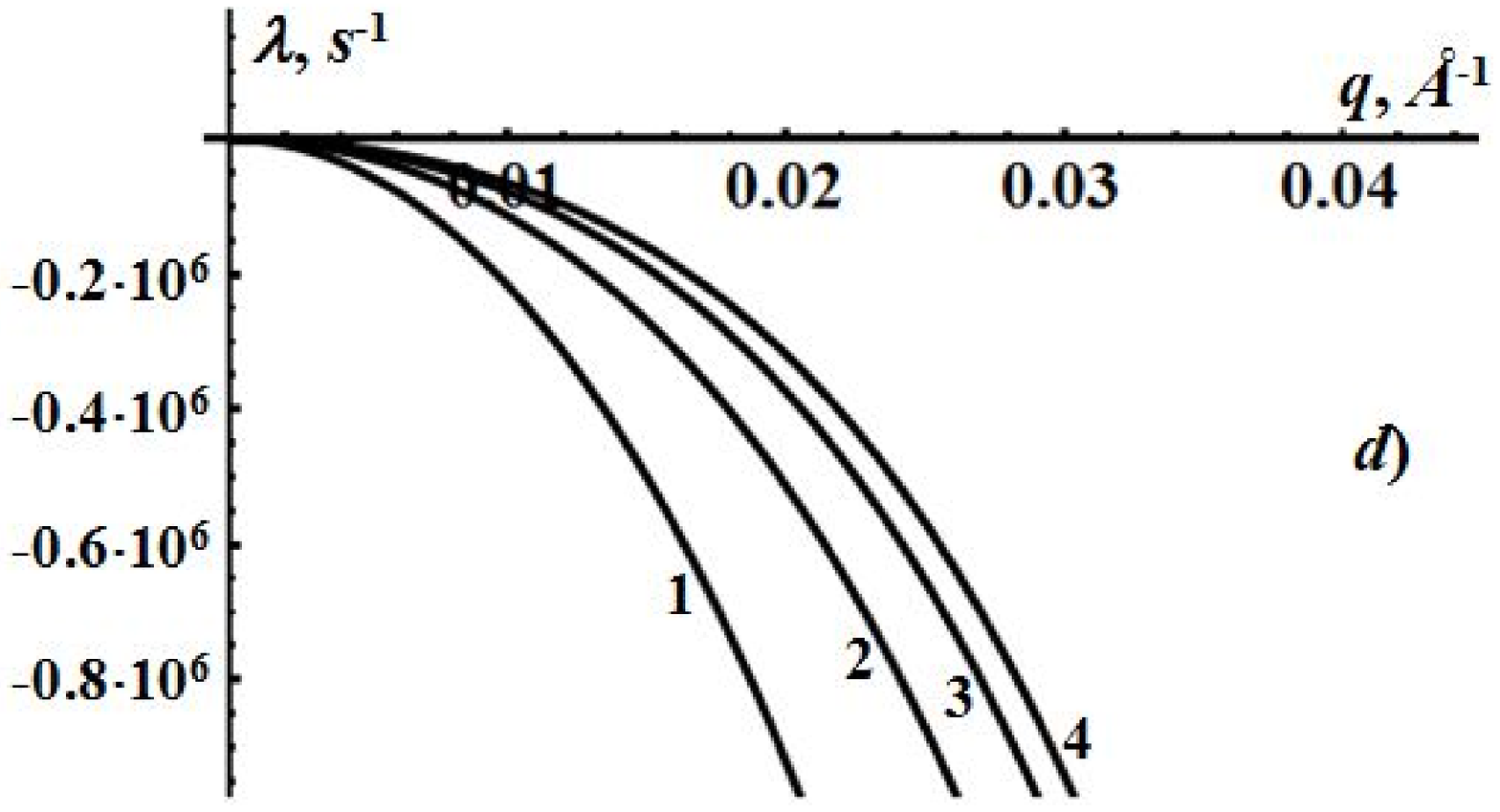}
\includegraphics[width=70mm]{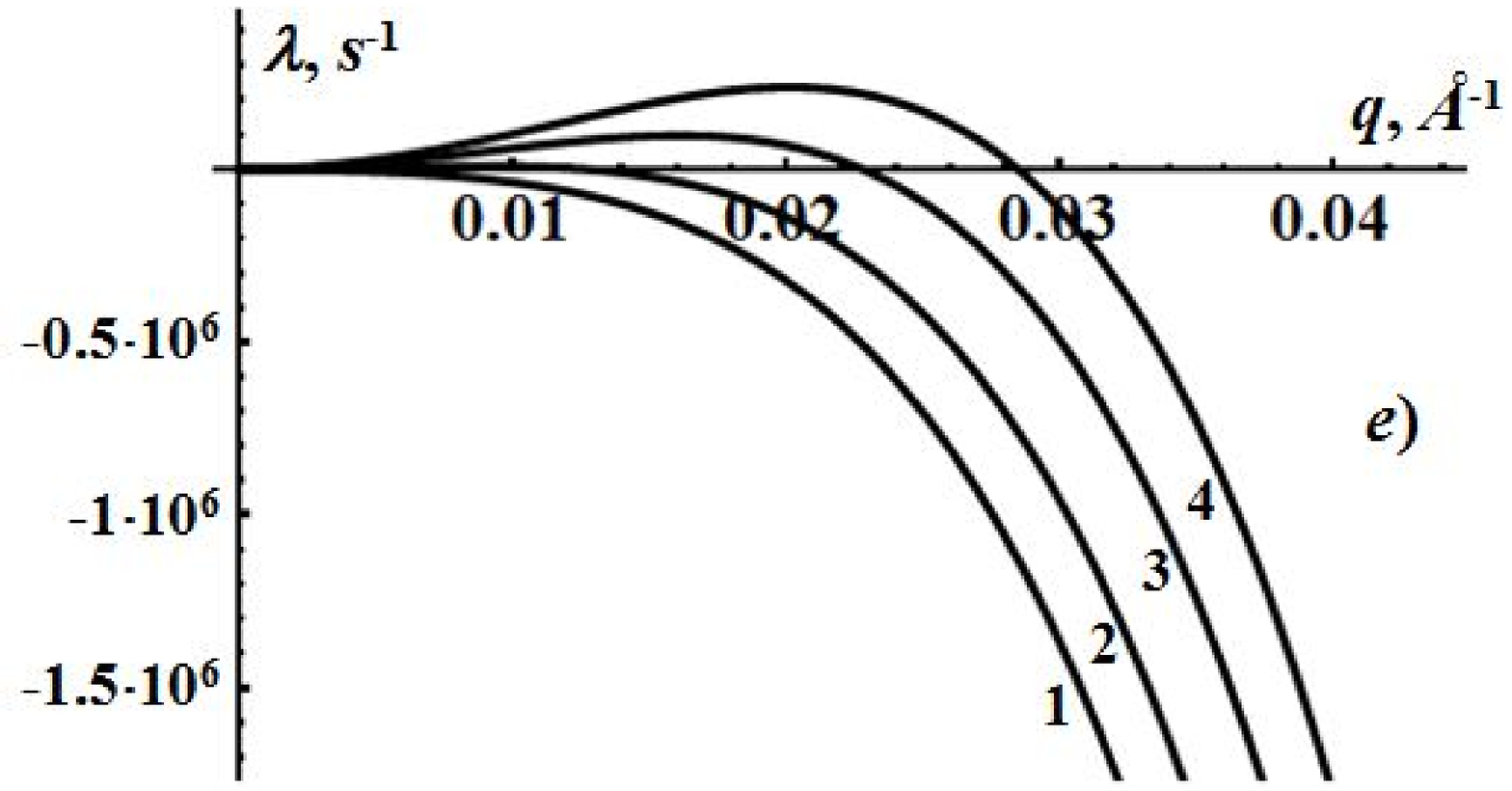}
\includegraphics[width=70mm]{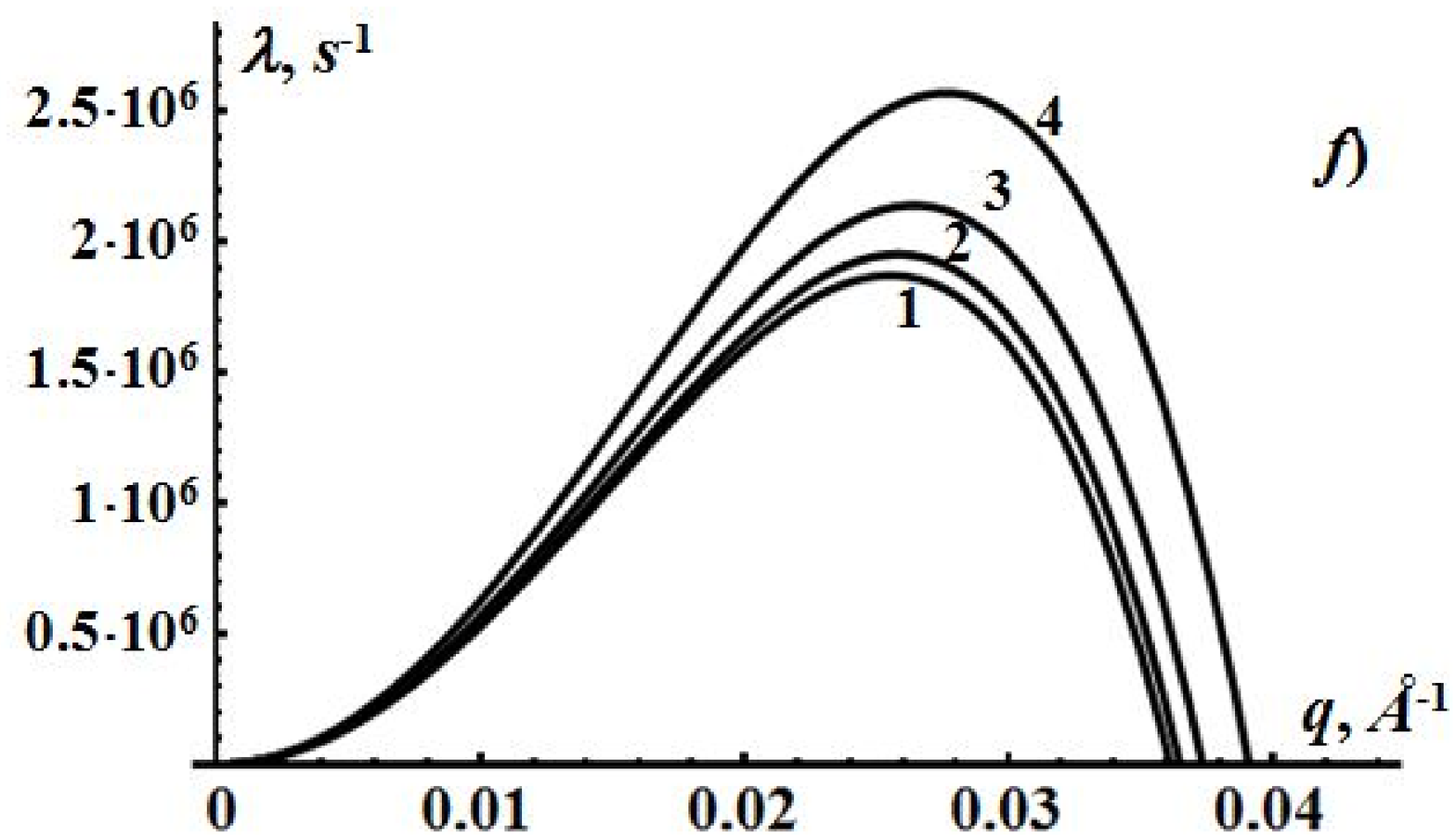}
\end{multicols}
\end{center}
\vspace{-3mm}
        \caption{The dependence of the increment of defect-deformation instability on the wave vector at concentration of the adsorbed atoms of
 $N_{0l}^ + = 2\times10^{12}$~cm$^{-2}$ at various values of the concentration of donors:
1~--- $ N_{0\textrm{Si}} = 10^8$~cm$^{-2}$; 2~--- $ N_{0\textrm{Si}} = 5\times10^{12}$~cm$^{-2}$; 3~--- $ N_{0\textrm{Si}} = 2\times10^{13}$~cm$^{-2}$; 4~---
$N_{0\textrm{Si}} = 10^{14}$~cm$^{-2}$; ({a})~$T = 70$~K; ({b}) $T = 150$~K; ({c})
$T = 250$~K; ({d}) $T = 300$~K; ({e}) $T = 500$~K; (f) $T = 800$~K.}
        \label{fig2}
    \end{figure}

\begin{figure}[!b]
 \begin{center}
\begin{multicols}{2}
\includegraphics[width=70mm]{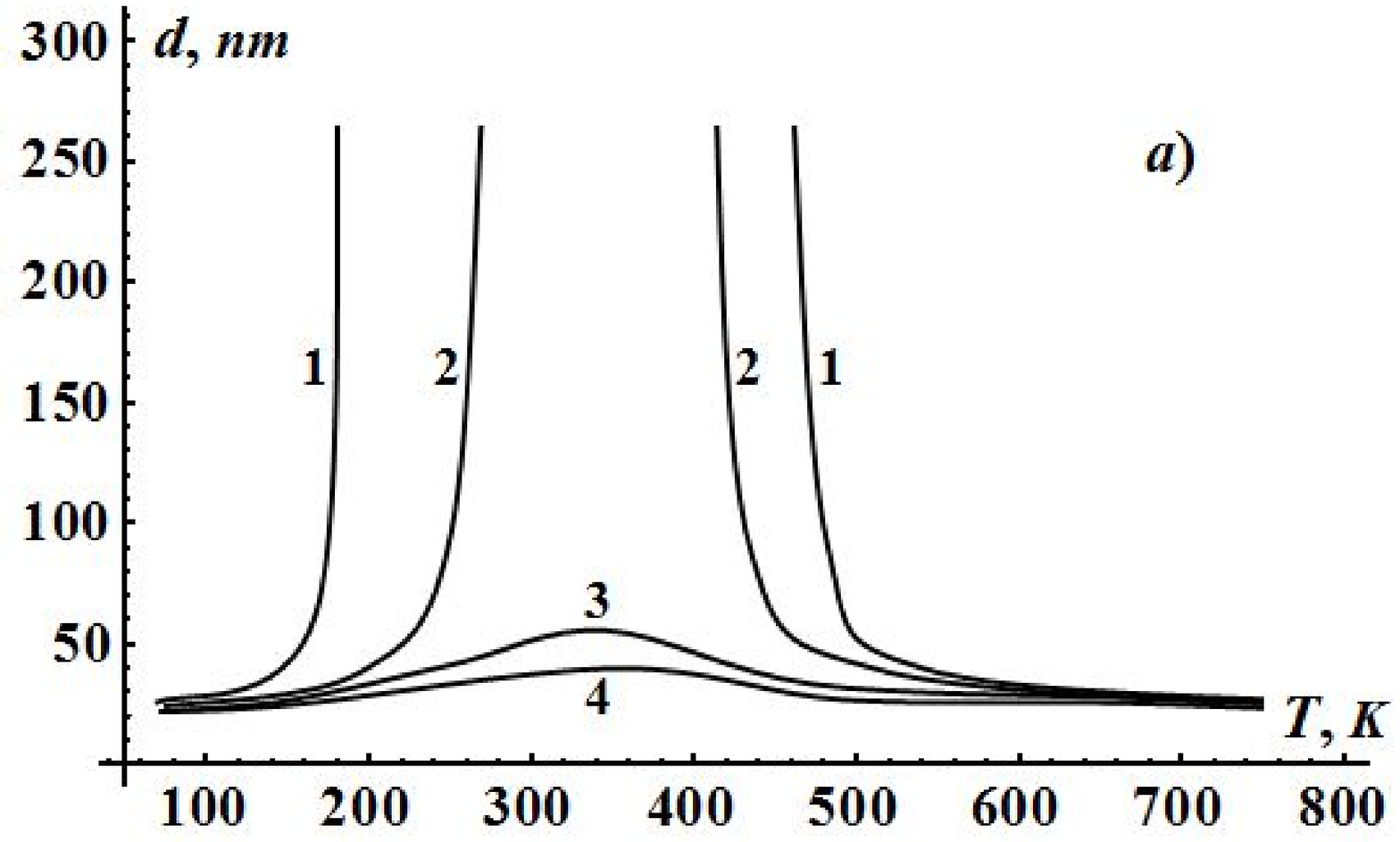}
\includegraphics[width=70mm]{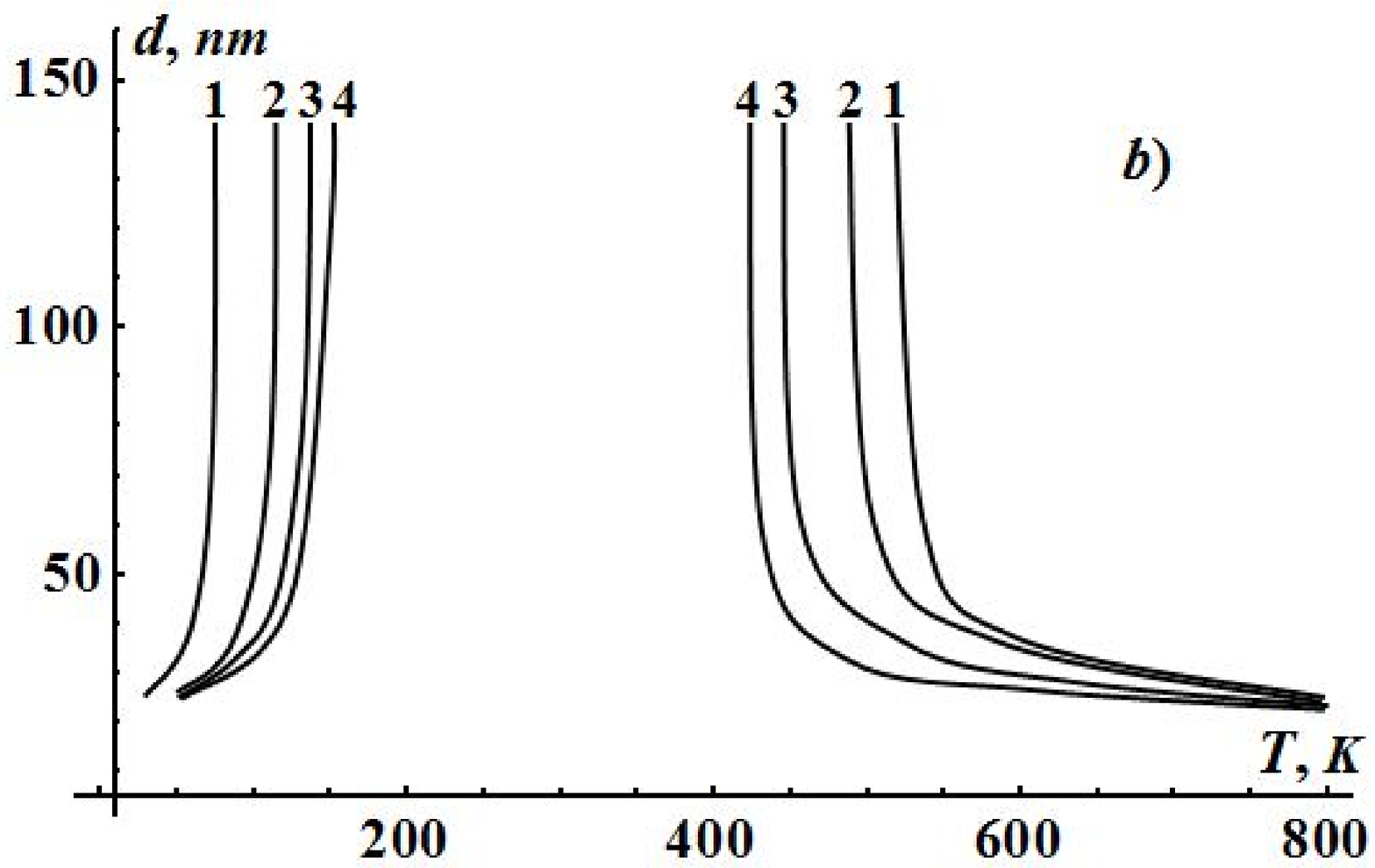}
\end{multicols}
\end{center}
\vspace{-4mm}
        \caption{The dependence of the period of the surface defect-deformation structure on temperature at various values of the concentration of donors:
1~--- $ N_{0\textrm{Si}} = 10^8$~cm$^{-2}$; 2~--- $ N_{0\textrm{Si}} = 5\times10^{12}$~cm$^{-2}$; 3~--- $ N_{0\textrm{Si}} = 2\times10^{13}$~cm$^{-2}$; 4~---
$ N_{0\textrm{Si}} = 10^{14}$~cm$^{-2}$; ({a}) $N_{0l}^ + = 5\times 10^{12}$~cm$^{-2}$; ({b}) $N_{0l}^ + = 2\times 10^{12}$~cm$^{-2}$.}
        \label{fig3}
    \end{figure}

The results of calculation of the dependence of the increment of defect-deformation instability on the module of the wave vector are given in figures~\ref{fig1} and~\ref{fig2} at various values of the concentration of silicon impurities $N_{0\textrm{Si}}$  and various values of temperature at average concentration of the Ga adatoms  $N_{0l}^ +   = {5} \times {\rm 10}^{{\rm 12}}$~cm$^{- 2}$ (figure~\ref{fig1}) and   $N_{0l}^ +   = {2} \times {\rm 10}^{{\rm 12}}$~cm$^{- 2}$ (figure~\ref{fig2}), which are generated under the action of laser radiation and are defined by its power. Such a dependence has a maximum which is shifted towards great values of the module of the wave vector with an increasing concentration of donors. The value of $q_{\textrm{max}}$ at which the increment of defect-deformation instability $\lambda$  has a maximum, defines the period of the dominating structure $d = {{2\pi }}/{{q_{\max } }}$  (figure~\ref{fig3}). The negative values of the increment of defect-deformation instability $\lambda$  mean the attenuation of SAW and, respectively, the impossibility of the formation of periodic defect-deformation structures. Except the structure dominating in the linear approximation with an increment  $\lambda_{\textrm{max}}(q_{\textrm{max}})$, the whole continuum of structures with  $\lambda > 0$  intensifies (figure~\ref{fig1} and figure~\ref{fig2}). At an exit to the stationary regime, the increase in amplitude of the dominant structure (\ref{eq4}), (\ref{eq7})--(\ref{eq9}) will reach saturation due to elastic nonlinearity, but amplitudes with a smaller increment will continue to increase \cite{Eme99}. Thus, the spectrum of the generated modes can extend. However, in  \cite{Eme99} it is shown that in solids with defects which are the centers of tension ($\theta  > 0$), only single-mode generation with $q_{\textrm{max}}$ is most often realized.

As seen from figure~\ref{fig1}, at concentration of the adatoms $N_{0l}^ +   = {\rm 5} \times {\rm 10}^{{\rm 12}}$~cm$^{ - 2}$, the formation of periodic structures is possible in a wide temperature range. In particular, at low temperatures $T < 180$~K [figure~\ref{fig1}~(a),~(b) and figure~\ref{fig3}~(a)] and high temperatures $T > 460$~K [figure~\ref{fig1}~(e),~(f) and figure~\ref{fig3}~(a)] at this intensity of laser irradiation, the formation of the surface periodic structures occurs at any degree of a doping of GaAs semiconductor matrix. At a significant doping of GaAs semiconductor by donor impurities of silicon ($N_{0\textrm{Si}} > 10^{13}$~cm$^{-2}$), the formation of periodic defect structures is possible at any temperature [figure~\ref{fig1} and figure~\ref{fig3}~(a)].

At a decrease of intensity of laser irradiation, the concentration of Ga adatoms decreases. In this case, the temperature ranges in which the formation of the surface superlattice is possible are narrowed. In particular, at concentration of adatoms
$N_{0l}^ +   = {\rm 2} \times {\rm 10}^{{\rm 12}}$~cm$^{ - 2}$, the formation of periodic structures in the temperature range $150~\textrm{K} < T < 420~\textrm{K}$ is impossible at any degree of a doping of GaAs semiconductor [figure~\ref{fig2} and figure~\ref{fig3}~(b)], while in undoped GaAs (curves~1), the formation of self-assembled defect structures can only occur at temperature $T < 80$~K [figure~\ref{fig2}~(a) and figure~\ref{fig3}~(b)] and $T > 520$~K [figure~\ref{fig2}~(f)  and figure~\ref{fig3}~(b)].

At an increase of concentration of donors, the acoustoelectronic effects lead to an increase of the value of the increment of defect-deformation instability. Thus, in a semiconductor with a larger degree of doping by donor impurities, the processes of the formation of nanometer periodic structures should occur quicker.

The dependence of the period of the surface superlattice of GaAs adatoms on temperature is presented in figure~\ref{fig3}. Such a dependence is of a nonmonotonic character.

At high intensity of laser irradiation ($N_{0l}^ +   = {\rm 5} \times {\rm 10}^{{\rm 12}}$~cm$^{ - 2}$) and at high concentration of donors [figure~\ref{fig3}~(a), curves~3 and 4], the greatest value of the period is observed at temperature $T=(320-380)$~K.

At  $T \to 0$ or  $T \to \infty$, the period of defect-deformation structure decreases and approaches the value  $2\pi l_\textrm{d}$. At a decrease of the degree of a doping (figure~\ref{fig3}, curves~1 and 2), there are two critical temperatures $T_{\textrm{c},\textrm{min}}$ and $T_{\textrm{c},\textrm{max}}$, for which only at $T < T_{\textrm{c},\textrm{min}}$ and $T > T_{\textrm{c},\textrm{max}}$, the formation of the surface superlattice is possible. At $T\rightarrow T_{\textrm{c},\textrm{min}}$ and $T\rightarrow T_{\textrm{c},\textrm{max}}$, the superlattice period increases monotonously and strives to infinity. At a decrease of intensity of laser irradiation (at a decrease of concentration of the adsorbed atoms), the temperature $T_{\textrm{c},\textrm{min}}$ decreases, and the temperature $T_{\textrm{c},\textrm{max}}$ increases [figure~\ref{fig3}~(b)].

Our results are qualitatively consistent with the experimental results presented in works \cite{Mac02, Vin04}. In particular, in \cite{Mac02} it is shown that
 the formation of nanoclusters of gallium is observed on the laser illuminated part of the end surface of the quartz light guide at molecular beam epitaxy of Ga at a temperature of 100~K. At $T = 300$~K, the effect of self-organization of nanoclusters is not observed. In work \cite{Vin04} it is shown that the formation of nanoclusters at room temperature on the surface of strongly alloyed semiconductor substrate of GaAs ($\sim10^{18}$~cm$^{-3}$)  is possible if the intensity of the laser radiation exceeds the defined critical value.

It should be noted that at a considerable concentration of adatoms, the non-linear effects that can slightly expand the temperature range at which the formation of nanoclusters is possible \cite{Pel09, Pel15}, are essential.

\section{Conclusions}

\begin{enumerate}
\item	The theory of nucleation of nanoclusters on ${n}$-GaAs surface under the action of laser irradiation which considers the doping degree of the semiconductor having donor impurities is developed.

\item	Within this theory, the temperature ranges of the formation of periodic defect structures on ${n}$-GaAs surface depending on the concentration of donors and on the intensity of laser radiation (concentration of the adsorbed atoms) are established.

\item It is shown that an increase of the degree of  doping of a semiconductor expands temperature ranges within which the formation of periodic defect structures is possible.

\item	The regularities of the effect of temperature at various values of the concentration of donors on the period of the surface defect-deformation structures of the adatoms in GaAs semiconductor are determined.
\end{enumerate}

\clearpage

\ukrainianpart

\title{Температурні  режими  формування  нанометрової періодичної структури адсорбованих атомів \\ у напівпровіднику  GaAs під дією лазерного опромінення}
\author[Р.М.~Пелещак, О.В.~Кузик, \ldots]{Р.М.~Пелещак, О.В.~Кузик, О.О.~Даньків}
\address{Дрогобицький державний педагогічний університет імені Івана Франка, \\
вул. Івана Франка, 24, 82100 Дрогобич, Україна
}

\makeukrtitle

\begin{abstract}
\tolerance=3000%
Розвинуто теорію нуклеації нанорозмірних структур адсорбованих атомів (адатомів), яка відбувається в результаті самоузгодженої взаємодії адатомів з поверхневою акустичною хвилею та електронною підсистемою. Досліджено температурні режими формування нанокластерів на поверхні ${n}$-GaAs під дією лазерного опромінення. Запропонована модель дозволяє вибрати оптимальні технологічні параметри (температуру, ступінь легування, інтенсивність лазерного опромінення) для формування поверхневих періодичних дефектно-деформаційних структур під дією лазерного опромінення.
\keywords нанокластер, температура, дифузія, деформація

\end{abstract}

\end{document}